\documentclass[preprint,aps,12pt,showpacs,nofootinbib,tightenlines]{revtex4}
\usepackage{amsmath}
\usepackage{amssymb}
\usepackage{epsfig}
\usepackage{graphicx}
\newcommand{\psl}{ P \hspace{-2.4truemm}/}
\newcommand{\epsl}{\epsilon \hspace{-1.8truemm}/\,  }
\def\be{\begin{eqnarray}}
\def\en{\end{eqnarray}}
\def\non{\nonumber\\}

\def\ra{\rangle}

\def\sl{\!\!\!\slash}
\def\prd{{Phys. Rev. D}~}
\def\prl{{ Phys. Rev. Lett.}~}
\def\plb{{ Phys. Lett. B}~}

\def\epjc{{ Eur. Phys. J. C}~}
\newcommand{\acp}{{\cal A}_{CP}}
\begin{document}
\title{Branching Ratio and CP Asymmetry of $B_s \to K^*_0(1430)\rho(\omega,\phi)$ Decays in the PQCD Approach}
\author{Zhi-Qing Zhang
\footnote{Electronic address: zhangzhiqing@haut.edu.cn} } 
\affiliation{\it \small  Department of Physics, Henan University of
Technology, Zhengzhou, Henan 450052, P.R.China } 
\date{\today}
\begin{abstract}
In the two-quark model
supposition for $K_0^{*}(1430)$, which can be viewed as either the first excited state (scenario I) or the lowest lying state (scenario II),
the branching ratios and the direct CP-violating
asymmetries for decays $\bar B_s^0\to K^{*0}_0(1430)\phi, K^{*0}_0(1430)\omega, K^{*0}_0(1430)\rho^0, K^{*+}_0(1430)\rho^-$
are studied by employing the perturbative QCD factorization approach. We find the following results: (a) Enhanced by
the color allowed tree amplitude with large Wilson coefficients $a_1=C_2+C_1/3$, the branching ratio of
$\bar B_s^0\to K^{*+}_0(1430)\rho^-$ is much larger than those of the other three decays and arrives at
$(3.4^{+0.8}_{-0.7})\times 10^{-5}$ in scenario I, even $10^{-4}$ order in scenario II, and its direct CP
violating asymmetry is the smallest, around $10\%$, so this channel might be measurable in the current LHC-b
experiments, where a large number (about $10^{12}$) of $B$ mesons will be produced per year. This high statistics will make the
measurement possible. (b) For the decay modes $\bar B^0_s\to K^{*0}_0(1430)\omega, K^{*0}_0(1430)\rho^0$, their direct CP-violating
asymmetries are large, but it might be difficult to measure them, because their branching ratios are small and less than
(or near) $10^{-6}$ in both scenarios. For example, in scenario I, these values are ${\cal B}(\bar B_s^0\to
K^*_0(1430)\omega)=(8.2^{+1.8}_{-1.7})\times 10^{-7},
{\cal B}(\bar B_s^0\to K^*_0(1430)\rho^0)=(9.9^{+2.1}_{-2.0})\times 10^{-7},
\acp^{dir}(\bar B^0_s\to K^{*0}_0(1430)\omega)=-24.1^{+2.8}_{-2.5},
\acp^{dir}(\bar B^0_s\to K^{*0}_0(1430)\rho^0)=26.6^{+2.5}_{-2.5}.$ (c) For the decay $\bar B^0_s\to K^*_0(1430)\phi$,
the predicted branching ratios
are also small and a few times $10^{-7}$ in both scenarios; there is no tree contribution at the
leading order, so its direct CP-violating asymmetry is naturally zero.
\end{abstract}

\vspace{1cm}

\maketitle


\section{Introduction}\label{intro}
Along with many scalar mesons found in experiments, more and more efforts have been made to study the scalar meson spectrum
theoretically \cite{nato,jaffe,jwei,baru,celenza,stro,close1}. Today, it is still a difficult but interesting topic. Our most important
task is to uncover the
mysterious structures of the scalar mesons.  There are two typical schemes for their classification \cite{nato,jaffe}. Scenario I: the nonet mesons below
1 GeV, including $f_0(600), f_0(980), K^*_0(800)$, and $a_0(980)$, are
usually viewed as the lowest lying $q\bar q$ states, while the nonet
ones near 1.5 GeV, including $f_0(1370), f_0(1500)/f_0(1700),
K^*_0(1430)$, and $a_0(1450)$, are suggested as the first excited
states. In scenario II, the nonet mesons near 1.5 GeV are
treated as $q\bar q$ ground states, while the nonet mesons below 1
GeV are exotic states beyond the quark model, such as four-quark
bound states.

In order to
uncover the inner structures of these scalar mesons, many factorization approaches are also used to research
the $B$ meson decay modes with a final state scalar meson, such as the generalized factorization approach
\cite{GMM}, QCD factorization approach
\cite{CYf0K,ccysp,ccysv}, and perturbative QCD (PQCD) approach
\cite{zqzhang1,zqzhang2,zqzhang3,zqzhang4,zqzhang5}.
On the experimental side, along with
the running of the Large Hadron Collider beauty (LHC-b) experiments, some of $B_s$ decays with a
scalar meson in the final state might be observed in the current \cite{lhc1,lhc2}.
In order to make precise measurements of rare decay rates and CP violating observables in the $B$-meson systems, the LHC-b
detector is designed to exploit the large number of $b$-hadrons produced. LHC-b will produce up to $10^{12}$ $b\bar b$
pairs per year $(10^7 s)$. Furthermore, it can reconstruct a $B$-decay
vertex with very good resolution, which is essential for studying the rapidly oscillating $B_s$ mesons.
In a word, $B_s$ decays with a scalar
in the final state can also serve as an
ideal platform to probe the natures of these scalar mesons. So the
studies of these decay modes for $B_s$ are necessary in the next a
few years.

Here $K^*_0(1430)$ can be treated as a $q\bar q$
state in both scenario I and scenario II, it is easy to make quantitative predictions in the two-quark  model
supposition, so we would like to use the PQCD approach to calculate the branching ratios
and the CP-violating
asymmetries for decays $\bar B_s^0\to K^{*0}_0(1430)\phi, K^{*0}_0(1430)\omega, K^{*0}_0(1430)\rho^0,
K^{*+}_0(1430)\rho^-$ in two scenarios.
In the following, $K^*_0(1430)$ is denoted as $K^*_0$ in some places for convenience.
The layout of this paper is as follows. In Sec. \ref{proper}, the decay constants
and light-cone distribution amplitudes of relevant mesons are introduced.
In Sec. \ref{results}, we then analyze these decay channels using the PQCD approach.
The numerical results and the discussions are given
in section \ref{numer}. The conclusions are presented in the final part.

\section{decay constants and distribution amplitudes }\label{proper}

In general, the $B_s$ meson is treated as a heavy-light system, and its
Lorentz structure can be written as\cite{grozin,kawa} \be
\Phi_{B_s}=\frac{1}{\sqrt{2N_c}}(\psl_{B_s}+M_{B_s})\gamma_5\phi_{B_s}(k_1).\label{bmeson}
\en
The contribution of $\bar \phi_{B_s}$ is numerically small
\cite{caidianlv} and has been neglected. For the distribution
amplitude $\phi_{B_s}(x,b)$ in Eq.(\ref{bmeson}), we adopt the following model:
\be
\phi_{B_s}(x,b)=N_{B_s}x^2(1-x)^2\exp[-\frac{M^2_{B_s}x^2}{2\omega^2_{b_s}}-\frac{1}{2}(\omega_{b_s}b)^2],
\en
where $\omega_{b_s}$ is a free parameter, we take
$\omega_{b_s}=0.5\pm0.05$ GeV in numerical calculations, and
$N_{B_s}=63.67$ is the normalization factor for $\omega_{b_s}=0.5$.

In  the two-quark picture, the vector decay constant $f_{K^*_0}$ and the
scalar decay constant $\bar {f}_{K^*_0}$ for the scalar meson
$K^*_0$ can  be defined as \be \langle K^*_0(p)|\bar q_2\gamma_\mu
q_1|0\ra&=&f_{K^*_0}p_\mu, \en \be \langle K^*_0(p)|\bar
q_2q_1|0\ra=m_{K^*_0}\bar {f}_{K^*_0}, \label{fbar} \en where
$m_{K^*_0}(p)$ is the mass (momentum) of the scalar meson $K^*_0$.
The relation between $f_{K^*_0}$ and $\bar f_{K^*_0}$ is \be
\frac{m_{{K^*_0}}}{m_2(\mu)-m_1(\mu)}f_{{K^*_0}}=\bar f_{{K^*_0}},
\en where $m_{1,2}$ are the running current quark masses. For the
scalar meson $K^*_0(1430)$, $f_{K^*_0}$ will get a very small value
after the $SU(3)$ symmetry breaking is considered. The light-cone
distribution amplitudes for the  scalar meson $K^*_0(1430)$
can be written as \be \langle K^*_0(p)|\bar q_1(z)_l
q_2(0)_j|0\rangle &=&\frac{1}{\sqrt{2N_c}}\int^1_0dx \; e^{ixp\cdot
z}\non && \times \{ p\sl\Phi_{K^*_0}(x)
+m_{K^*_0}\Phi^S_{K^*_0}(x)+m_{K^*_0}(n\sl_+n\sl_--1)\Phi^{T}_{K^*_0}(x)\}_{jl}.\quad\quad\label{LCDA}
\en Here $n_+$ and $n_-$ are lightlike vectors:
$n_+=(1,0,0_T),n_-=(0,1,0_T)$, and $n_+$ is parallel with the moving
direction of the scalar meson. The normalization can be related to
the decay constants: \be \int^1_0 dx\Phi_{K^*_0}(x)=\int^1_0
dx\Phi^{T}_{K^*_0}(x)=0,\,\,\,\,\,\,\,\int^1_0
dx\Phi^{S}_{K^*_0}(x)=\frac{\bar f_{K^*_0}}{2\sqrt{2N_c}}\;. \en The
twist-2 light-cone distribution amplitude $\Phi_{K^*_0}$ can be expanded in the Gegenbauer
polynomials: \be \Phi_{K^*_0}(x,\mu)&=&\frac{\bar
f_{K^*_0}(\mu)}{2\sqrt{2N_c}}6x(1-x)\left[B_0(\mu)+\sum_{m=1}^\infty
B_m(\mu)C^{3/2}_m(2x-1)\right], \en where the decay constants and
the Gegenbauer moments $B_1,B_3$ of distribution amplitudes for
$K^*_0(1430)$ have been calculated in the QCD sum rules\cite{ccysp}.
These values are all scale dependent and specified below: \be
{\rm scenario I:} B_1&=&0.58\pm0.07, B_3=-1.2\pm0.08, \bar f_{K^*_0}=-(300\pm30){\rm MeV},\\
{\rm scenario II:}B_1&=&-0.57\pm0.13, B_3=-0.42\pm0.22, \bar f_{K^*_0}=(445\pm50){\rm MeV},\quad
\en
which are taken by fixing the scale at 1GeV.

As for the twist-3 distribution amplitudes $\Phi_{K^*_0}^S$ and $\Phi_{K^*_0}^T$, we adopt the asymptotic form:
\be
\Phi^S_{K^*_00}&=& \frac{1}{2\sqrt {2N_c}}\bar f_{K^*_0},\,\,\,\,\,\,\,\Phi_{K^*_0}^T=
\frac{1}{2\sqrt {2N_c}}\bar f_{K^*_0}(1-2x).
\en

The distribution amplitudes up to twist-3 of the vector mesons are
\be
\langle V(P,\epsilon^*_L)|\bar q_{2\beta}(z)q_{1\alpha}(0)|0\rangle=\frac{1}{2N_C}\int^1_0dxe^{ixP\cdot z}[M_V\epsl^*_L\Phi_V(x)
+\epsl_L^*\psl\Phi_V^t(x)+M_V\Phi^s_V(x)]_{\alpha\beta},\quad
\en
for longitudinal polarization. The distribution amplitudes
can be parametrized as
\be
\Phi_V(x)&=&\frac{2f_V}{\sqrt{2N_C}}[1+a^{\|}_2C^{\frac{3}{2}}_2(2x-1)],\\
\Phi_V^t(x)&=&\frac{3f^T_V}{2\sqrt{2N_C}}(2x-1)^2,\quad \phi_V^s(x)=-\frac{3f^T_V}{2\sqrt{2N_C}}(2x-1),
\en
where the decay constant $f_V$ \cite{yao} and the transverse decay constant $f^T_V$ \cite{pball} are given as the following values:
\be
f_\rho&=&209\pm2 {\rm MeV}, f_\omega=195\pm3 {\rm MeV}, f_\phi=231\pm4 {\rm MeV}, \\
f^T_\rho&=&165\pm9 {\rm MeV}, f^T_\omega=151\pm9 {\rm MeV}, f^T_\phi=186\pm9 {\rm MeV}.
\en
Here the Gegenbauer polynomial is defined as $C^{\frac{3}{2}}_2(t)=\frac{3}{2}(5t^2-1)$. For the Gegenbauer moments, we quote the
numerical results as \cite{pball1}:
\be
a^{\|}_{2\rho}=a^{\|}_{2\omega}=0.15\pm0.07, a^{\|}_{2\phi}=0.18\pm0.08.
\en


\section{ the perturbative QCD  calculation} \label{results}

Under the two-quark model for the scalar meson $K^*_0$ supposition,
the decay amplitude for $\bar B^0_s\to VK^*_0$, where $V$ represents $\rho, \omega, \phi$,
 can be conceptually written as the convolution,
\be
{\cal A}(\bar B^0_s \to V K^*_0)\sim \int\!\! d^4k_1
d^4k_2 d^4k_3\ \mathrm{Tr} \left [ C(t) \Phi_{B_s}(k_1) \Phi_{V}(k_2)
\Phi_{K^*_0}(k_3) H(k_1,k_2,k_3, t) \right ], \label{eq:con1}
\en
where $k_i$'s are momenta of the antiquarks included in each meson, and
$\mathrm{Tr}$ denotes the trace over Dirac and color indices. $C(t)$
is the Wilson coefficient which results from the radiative
corrections at short distance. In the above convolution, $C(t)$
includes the harder dynamics at larger scale than the $M_B$ scale and
describes the evolution of local $4$-Fermi operators from $m_W$ (the
$W$ boson mass) down to $t\sim\mathcal{O}(\sqrt{\bar{\Lambda} M_{B_s}})$
scale, where $\bar{\Lambda}\equiv M_{B_s} -m_b$. The function
$H(k_1,k_2,k_3,t)$ describes the four-quark operator and the
spectator quark connected by
 a hard gluon whose $q^2$ is in the order
of $\bar{\Lambda} M_{B_s}$, and includes the
$\mathcal{O}(\sqrt{\bar{\Lambda} M_{B_s}})$ hard dynamics. Therefore,
this hard part $H$ can be perturbatively calculated. The functions
$\Phi_{(V, K^*_0)}$ are the wave functions of the vector meson $V$ and
the scalar meson $K^*_0$, respectively.

Since the $b$ quark is rather heavy, we consider the $B_s$ meson at rest
for simplicity. It is convenient to use the light-cone coordinate $(p^+,
p^-, {\bf p}_T)$ to describe the meson's momenta, \be p^\pm =
\frac{1}{\sqrt{2}} (p^0 \pm p^3), \quad {\rm and} \quad {\bf p}_T =
(p^1, p^2). \en Using these coordinates, the $B_s$ meson and the two
final state meson momenta can be written as \be P_{B_s} =
\frac{M_{B_s}}{\sqrt{2}} (1,1,{\bf 0}_T), \quad P_{2} =
\frac{M_{B_s}}{\sqrt{2}}(1-r^2_{K^*_0},r^2_V,{\bf 0}_T), \quad P_{3} =
\frac{M_{B_s}}{\sqrt{2}} (r^2_{K^*_0},1-r^2_V,{\bf 0}_T), \en respectively, where the ratio $r_{K^*_0(V)}=m_{K^*_0(V)}/M_{B_s}$, and
$m_{K^*_0(V)}$ is the scalar meson $K^*_0$ (the vector meson $V$) mass. Putting the antiquark momenta in $B_s$,
$V$, and $K^*_0$ mesons as $k_1$, $k_2$, and $k_3$, respectively, we can
choose
\be k_1 = (x_1 P_1^+,0,{\bf k}_{1T}), \quad k_2 = (x_2
P_2^+,0,{\bf k}_{2T}), \quad k_3 = (0, x_3 P_3^-,{\bf k}_{3T}). \en
For these considered decay channels, the integration over $k_1^-$,
$k_2^-$, and $k_3^+$ in Eq.(\ref{eq:con1}) will lead to
\be
 {\cal
A}(B_s \to V K^*_0) &\sim &\int\!\! d x_1 d x_2 d x_3 b_1 d b_1 b_2 d
b_2 b_3 d b_3 \non && \cdot \mathrm{Tr} \left [ C(t) \Phi_{B_s}(x_1,b_1)
\Phi_{V}(x_2,b_2) \Phi_{K^*_0}(x_3, b_3) H(x_i, b_i, t) S_t(x_i)\,
e^{-S(t)} \right ], \quad \label{eq:a2}
\en
where $b_i$ is the
conjugate space coordinate of $k_{iT}$, and $t$ is the largest
energy scale in function $H(x_i,b_i,t)$.
In order to smear the end-point singularity on $x_i$,
the jet function $S_t(x)$ \cite{li02}, which comes from the
resummation of the double logarithms $\ln^2x_i$, is used.
The last term $e^{-S(t)}$ in Eq.(\ref{eq:a2}) is the Sudakov form factor which suppresses
the soft dynamics effectively \cite{soft}.

 For the considered decays, the related weak effective
Hamiltonian $H_{eff}$ can be written as \cite{buras96}
\be
\label{eq:heff} {\cal H}_{eff} = \frac{G_{F}} {\sqrt{2}} \,
\left[\sum_{p=u,c}V_{pb} V_{pd}^* \left (C_1(\mu) O_1^p(\mu) +
C_2(\mu) O_2^p(\mu) \right) -V_{tb} V_{td}^*\sum_{i=3}^{10} C_{i}(\mu) \,O_i(\mu)
\right] .
\en
Here the Fermi constant $G_{F}=1.166 39\times
10^{-5} GeV^{-2}$ and the functions $Q_i (i=1,...,10)$ are the local four-quark operators. We specify below
the operators in ${\cal H}_{eff}$ for $b \to d$ transition: \be
\begin{array}{llllll}
O_1^{u} & = &  \bar d_\alpha\gamma^\mu L u_\beta\cdot \bar
u_\beta\gamma_\mu L b_\alpha\ , &O_2^{u} & = &\bar
d_\alpha\gamma^\mu L u_\alpha\cdot \bar
u_\beta\gamma_\mu L b_\beta\ , \\
O_3 & = & \bar d_\alpha\gamma^\mu L b_\alpha\cdot \sum_{q'}\bar
 q_\beta'\gamma_\mu L q_\beta'\ ,   &
O_4 & = & \bar d_\alpha\gamma^\mu L b_\beta\cdot \sum_{q'}\bar
q_\beta'\gamma_\mu L q_\alpha'\ , \\
O_5 & = & \bar d_\alpha\gamma^\mu L b_\alpha\cdot \sum_{q'}\bar
q_\beta'\gamma_\mu R q_\beta'\ ,   & O_6 & = & \bar
d_\alpha\gamma^\mu L b_\beta\cdot \sum_{q'}\bar
q_\beta'\gamma_\mu R q_\alpha'\ , \\
O_7 & = & \frac{3}{2}\bar d_\alpha\gamma^\mu L b_\alpha\cdot
\sum_{q'}e_{q'}\bar q_\beta'\gamma_\mu R q_\beta'\ ,   & O_8 & = &
\frac{3}{2}\bar d_\alpha\gamma^\mu L b_\beta\cdot
\sum_{q'}e_{q'}\bar q_\beta'\gamma_\mu R q_\alpha'\ , \\
O_9 & = & \frac{3}{2}\bar d_\alpha\gamma^\mu L b_\alpha\cdot
\sum_{q'}e_{q'}\bar q_\beta'\gamma_\mu L q_\beta'\ ,   & O_{10} & =
& \frac{3}{2}\bar d_\alpha\gamma^\mu L b_\beta\cdot
\sum_{q'}e_{q'}\bar q_\beta'\gamma_\mu L q_\alpha'\ ,
\label{eq:operators}
\end{array}
\en
where $\alpha$ and $\beta$ are
the $SU(3)$ color indices; $L$ and $R$ are the left- and
right-handed projection operators with $L=(1 - \gamma_5)$, $R= (1 +
\gamma_5)$. The sum over $q'$ runs over the quark fields that are
active at the scale $\mu=O(m_b)$, i.e., $(q'\epsilon\{u,d,s,c,b\})$.


\begin{figure}[t,b]
\vspace{-3cm} \centerline{\epsfxsize=16 cm \epsffile{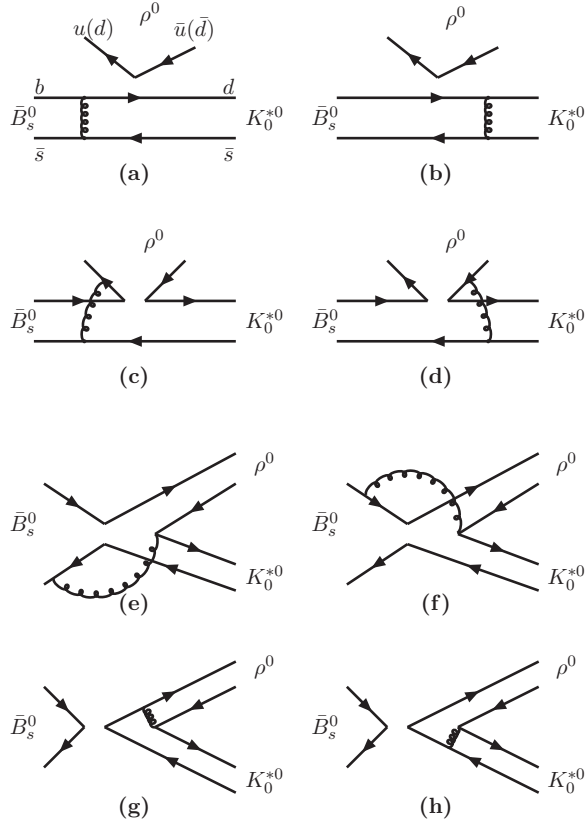}}
\vspace{-9cm} \caption{ Diagrams contributing to the decay $\bar
B_s^0\to \rho^0 K^{*0}_0(1430)$ .}
 \label{fig1}
\end{figure}

In Fig.~1, we give the leading order Feynman diagrams for the channel $\bar B_s^0\to \rho^0 K^{*0}_0(1430)$ as an example. The
Feynman diagrams for the other decays are similar and not given. The analytic
formulas of each considered decays are similar to those of $B\to f_0(980)K^*$ \cite{zqzhang4} and $B\to K^*_0(1430)\rho(\omega)$ \cite{zqzhang5}.
We just need to replace some corresponding wave functions, Wilson coefficients, and parameters. Here we do not show these formulas.

Combining the contributions from different diagrams, the total decay
amplitudes for these decays can be written as
\be
\sqrt{2}{\cal M}(K^{*0}_0\rho^0)&=&\xi_u\left[M_{eK^*_0}C_2
+F_{eK^*_0}a_2\right]-\xi_t\left[F_{eK^*_0}\left(-a_4+\frac{1}{2}(3C_7+C_8)
+\frac{5}{3}C_9+C_{10}\right)\right.\non &&\left.+M_{eK^*_0}(-\frac{C_3}{3}+\frac{C_9}{6}+\frac{3C_{10}}{2})
-(M^{P1}_{eK^*_0}+M^{P1}_{aK^*_0})(C_5-\frac{C_7}{2})+M^{P2}_{eK^*_0}\frac{3C_{8}}{2}
\right.\non &&\left.-M_{aK^*_0}(C_3-\frac{1}{2}C_9)-F_{aK^*_0}(a_4-\frac{1}{2}a_{10})-F^{P2}_{aK^*_0}(a_6-\frac{1}{2}a_8)\right],
\en
\be
\sqrt{2}{\cal M}(K^{*0}_0\omega)&=&\xi_u\left[M_{eK^*_0}C_2
+F_{eK^*_0}a_2\right]-\xi_t\left[F_{eK^*_0}\left(\frac{7C_3}{3}+\frac{5C_4}{3}+2a_5+\frac{a_7}{2}+\frac{C_9}{3}
-\frac{C_{10}}{3}\right)\right.\non &&\left.+M_{eK^*_0}(\frac{C_3}{3}+2C_4-\frac{C_9}{6}+\frac{C_{10}}{2})
+(M^{P1}_{eK^*_0}+M^{P1}_{aK^*_0})(C_5-\frac{C_7}{2})
\right.\non &&\left.+M^{P2}_{eK^*_0}(2C_6+\frac{C_{8}}{2})+M_{aK^*_0}(C_3-\frac{1}{2}C_9)+F_{aK^*_0}(a_4-\frac{1}{2}a_{10})\right.\non &&\left.
+F^{P2}_{aK^*_0}(a_6-\frac{1}{2}a_8)\right],
\en
\be
{\cal M}(K^{*+}_0\rho^-)&=&\xi_u\left[M_{eK^*_0}C_1
+F_{eK^*_0}a_1\right]-\xi_t\left[F_{eK^*_0}\left(a_4+a_{10}\right)+M_{eK^*_0}(C_3+C_9)\right.\non &&\left.
+M^{P1}_{eK^*_0}(C_5+C_7)
+M_{aK^*_0}(C_3-\frac{1}{2}C_9)+M^{P1}_{aK^*_0}(C_5-\frac{1}{2}C_7)\right.\non &&\left.+F_{aK^*_0}(a_4-\frac{1}{2}a_{10})
+F^{P2}_{aK^*_0}(a_6-\frac{1}{2}a_8)\right],\\
{\cal M}(K^{*0}_0\phi)&=&-\xi_t\left[F^{P2}_{e\phi}(a_6-\frac{a_8}{2})+M_{e\phi}(C_3-\frac{C_9}{2})
+(M^{P1}_{e\phi}+M^{P1}_{a\phi})(C_5-\frac{C_7}{2})\right.\non &&\left.
+M_{a\phi}(C_3-\frac{1}{2}C_9)+F_{a\phi}(a_4-\frac{1}{2}a_{10})+F_{a\phi}(a_6-\frac{1}{2}a_8)\right.\non &&\left.
+F_{eK^*_0}\left(a_3+a_5-\frac{1}{2}a_7-\frac{1}{2}a_{7}\right)+M_{eK^*_0}(C_4-\frac{1}{2}C_{10})\right.\non &&\left.
+M^{P2}_{eK^*_0}(C_6-\frac{1}{2}C_8)\right].
\en

The
combinations of the Wilson coefficients are defined as usual
\cite{zjxiao}:
 \be
a_{1}(\mu)&=&C_2(\mu)+\frac{C_1(\mu)}{3}, \quad
a_2(\mu)=C_1(\mu)+\frac{C_2(\mu)}{3},\non
a_i(\mu)&=&C_i(\mu)+\frac{C_{i+1}(\mu)}{3},\quad
i=3,5,7,9,\non
a_i(\mu)&=&C_i(\mu)+\frac{C_{i-1}(\mu)}{3},\quad
i=4, 6, 8, 10.\label{eq:aai} \en

\section{Numerical results and discussions} \label{numer}

We use the following input parameters in the numerical calculations \cite{pdg08}:
\be
f_{B_s}&=&230 MeV, M_{B_s}=5.37 GeV, M_W=80.41 GeV, \\
V_{ub}&=&|V_{ub}|e^{-i\gamma}=3.93\times10^{-3}e^{-i68^\circ}, V_{ud}=0.974, \\
 V_{td}&=&|V_{td}|e^{-i\beta}=8.1\times10^{-3}e^{-i21.6^\circ}, V_{tb}=1.0, \\
\alpha&=&100^\circ\pm20^\circ, \tau_{B_s}=1.470\times 10^{-12} s.
\en
Using the wave functions and the values of relevant input parameters, we find the numerical values
of the corresponding form factors $\bar B^0_s\to \phi, K^*_0(1430)$ at zero momentum transfer
\be
A^{\bar B^0_s\to \phi}_0(q^2=0)&=&0.29^{+0.05+0.01}_{-0.04-0.01},
\\F^{\bar B^0_s\to K^*_0}_0(q^2=0)&=&-0.30^{+0.03+0.01+0.01}_{-0.03-0.01-0.01}, \quad\mbox{ scenario I},
\\F^{\bar B^0_s\to K^*_0}_0(q^2=0)&=&0.56^{+0.05+0.03+0.04}_{-0.07-0.04-0.05}, \quad\;\;\;\mbox{ scenario II},
\en
where the uncertainties are from $\omega_{b_s}=0.5\pm0.05$ of $B_s$ and the Gegenbauer moment $a_{2\phi}=0.18\pm0.08$ of the vector
meson $\phi$ for $A^{\bar B^0_s\to \phi}$,
and from the decay constant,
the Gegenbauer moments $B_1$ and $B_3$ of the scalar meson $K^*_0$ for $F^{\bar B^0_s\to K^*_0}$. For the $\bar B_s\to\phi$ transition form
factor, its value is about $0.30$, which is favored by many model calculations \cite{wuyl,lucd,chenghy}, while a large value
$A^{\bar B_s\to\phi}_0=0.474$ is obtained by the light-cone sum-rule method \cite{pball1}. The discrepancy can be clarified by the
current LHC-b experiments.
As for the form factors $F^{\bar B^0_s\to K^*_0}_0$ in two scenarios, they are agree well with those given in \cite{lirh}.

In the $B_s$-rest frame, the decay rates of $\bar B^0_s\to K^*_0(1430)\rho(\omega,\phi)$ can be written as
\be
\Gamma=\frac{G_F^2}{32\pi m_{B_s}}|{\cal M}|^2(1-r^2_{K^*_0}),
\en
where ${\cal M}$ is the total decay amplitude of each
considered decay and $r_{K^*_0}$ is the mass ratio, both of which have been given  in  Sec. \ref{results}. The ${\cal M}$ can be rewritten as
\be
{\cal M}= V_{ub}V^*_{ud}T-V_{tb}V^*_{td}P=V_{ub}V^*_{ud}\left[1+ze^{i(\alpha+\delta)}\right] \label{ampde},
\en
where $\alpha$ is the Cabibbo-Kobayashi-Maskawa weak phase angle, and $\delta$ is the relative strong phase between
the tree and the penguin amplitudes, which are denoted as "T" and "P," respectively. The term $z$ describes the ratio of penguin to tree
contributions and is defined as
\be
z=\left|\frac{V_{tb}V^*_{td}}{V_{ub}V^*_{ud}}\right|\left|\frac{P}{T}\right|.
\en
From Eq.(\ref{ampde}), it is easy to write decay amplitude $\overline {\cal M}$ for the corresponding conjugated decay mode. So the CP-averaged
branching ratio for each considered decay is defined as
\be
{\cal B}=(|{\cal M}|^2+|\overline{\cal M}|^2)/2=|V_{ub}V^*_{ud}T|^2\left[1+2z\cos\alpha\cos\delta+z^2\right].\label{brann}
\en
\begin{figure}[t,b]
\begin{center}
\includegraphics[scale=0.7]{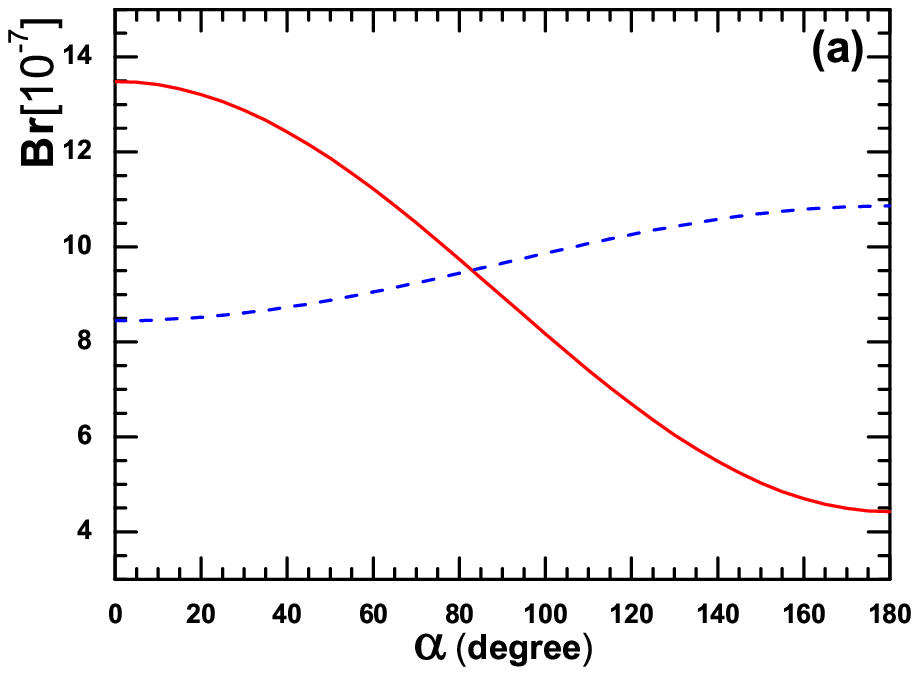}
\includegraphics[scale=0.7]{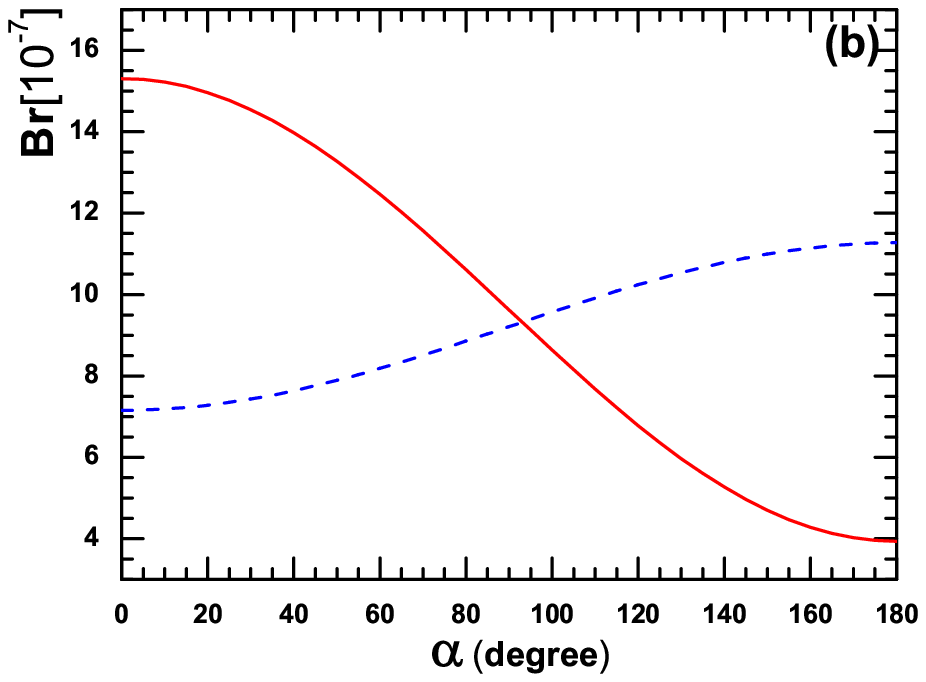}
\vspace{0.3cm} \caption{The dependence of the branching ratios for
$\bar B_s^0\to K^{*0}_0(1430)\omega$ (solid curve), $\bar B_s^0\to K^{*0}_0(1430)\rho^0$ (dashed curve) on the
Cabibbo-Kobayashi-Maskawa angle $\alpha$. The left (right) panel is plotted in scenario I (II).}\label{fig2}
\end{center}
\end{figure}
Using the input parameters and the wave functions as specified in this section and Sec. II, we can calculate
the branching ratios of the considered modes
\be
{\cal B}(\bar B_s^0\to
K^{*0}_0(1430)\phi)&=&(2.9^{+0.6+0.2+0.6}_{-0.5-0.1-0.5})\times 10^{-7}, \mbox{ scenario I},\\
{\cal B}(\bar B_s^0\to
K^{*0}_0(1430)\omega)&=&(8.2^{+1.7+0.0+0.6}_{-1.6-0.1-0.6})\times 10^{-7}, \mbox{ scenario I},\\
{\cal B}(\bar B_s^0\to
K^{*0}_0(1430)\rho^0)&=&(9.9^{+2.0+0.0+0.7}_{-1.9-0.1-0.7})\times 10^{-7}, \mbox{ scenario I},\\
{\cal B}(\bar B_s^0\to
K^{*+}_0(1430)\rho^-)&=&(3.4^{+0.7+0.3+0.3}_{-0.6-0.2-0.2})\times 10^{-5}, \mbox{ scenario I},\\
{\cal B}(\bar B_s^0\to
K^{*0}_0(1430)\phi)&=&(9.5^{+2.5+2.8+3.1}_{-1.7-1.9-1.4})\times 10^{-7}, \mbox{ scenario II},\\
{\cal B}(\bar B_s^0\to
K^{*0}_0(1430)\omega)&=&(8.6^{+2.1+0.6+2.2}_{-1.8-0.5-1.5})\times 10^{-7}, \mbox{ scenario II},\\
{\cal B}(\bar B_s^0\to
K^{*0}_0(1430)\rho^0)&=&(9.6^{+2.2+0.4+2.0}_{-2.0-0.4-2.1})\times 10^{-7}, \mbox{ scenario II},\\
{\cal B}(\bar B_s^0\to
K^{*+}_0(1430)\rho^-)&=&(10.8^{+2.5+1.2+1.9}_{-2.3-1.1-1.7})\times 10^{-5}, \mbox{ scenario II},
\end{eqnarray}
where the uncertainties are mainly from the decay constant,
the Gegenbauer moments $B_1$ and $B_3$ of the scalar meson $K^*_0$. From the results, one can find that the branching
ratios of $\bar B_s^0\to K^{*0}_0(1430)\phi, K^{*+}_0(1430)\rho^-$ in scenario II are about $3.2\sim3.3$ times larger than those in scenario I.
While for the decays $\bar B_s^0\to K^{*0}_0(1430)\omega(\rho^0)$, their branching ratios for two scenarios are very close to each other,
respectively. In these four decay channels, the branching ratio of $\bar B_s^0\to K^{*+}_0(1430)\rho^-$ is the largest one. This is not a surprise:
one can recall that the channel $\bar B_s^0\to K^{+}_0\rho^-$ also receives a large branching ratio, about
$(2.45^{+1.52}_{-1.29})\times10^{-5}$ predicted by the QCD factorization  approach \cite{Beneke} and about $(1.78^{+0.78}_{-0.59})
\times10^{-5}$ predicted by the PQCD approach \cite{lucd0}. Certainly, for the other three decays $\bar B_s^0\to K^{*0}_0(1430)\phi,
K^{*0}_0(1430)\omega(\rho^0)$,
their branch ratios have the same order with those of the decays $\bar B_s^0\to K^0_0\phi, K^0_0\omega(\rho^0)$, which are listed in Table I. It
is easy to get the conclusion that the branching ratios of the decays $\bar B_s^0\to K^*_0(1430)V$ are not far away from those of
$\bar B_s^0\to KV$, where $V$ represents $\rho, \omega, \phi$. The same conclusion is also obtained in Ref.\cite{zqzhang1}.
\begin{table}
\caption{Comparing the branching ratios  of $\bar B_s^0\to K^0_0\phi, K^0_0\omega, K^0_0\rho^0, K^+_0\rho^-$ predicted in \cite{Beneke}
and those of $\bar B_s^0\to K^{*0}_0(1430)\phi, K^{*0}_0(1430)\omega, K^{*0}_0(1430)\rho^0, K^{*+}_0(1430)\rho^-$
predicted in this work in scenario I . }\label{para}
\begin{center}
\begin{tabular}{c|c}
\hline \hline
 Mode& \cal{Br}($\times 10^{-6}$)\\
 \hline
$\bar B_s^0\to K^{*0}_0(1430)\phi$ & $0.29^{+0.06+0.02+0.06}_{-0.05-0.01-0.05}$\\
$\bar B_s^0\to K^0_0\phi$& $0.27^{+0.09+0.28+0.09+0.67}_{-0.08-0.14-0.06-0.18}$\\
\hline
$\bar B_s^0\to K^{*0}_0(1430)\omega$ &$0.82^{+0.17+0.00+0.06}_{-0.16-0.01-0.06}$ \\
$\bar B_s^0\to K^0_0\omega$ &$0.51^{+0.20+0.15+0.68+0.40}_{-0.18-0.11-0.23-0.25}$ \\
\hline
$\bar B_s^0\to K^{*0}_0(1430)\rho^0$& $0.99^{+0.02+0.00+0.07}_{-0.19-0.01-0.07}$\\
$\bar B_s^0\to K^0_0\rho^0$& $0.61^{+0.33+0.21+1.06+0.56}_{-0.26-0.15-0.38-0.36}$\\
\hline
$\bar B_s^0\to K^{*+}_0(1430)\rho^-$& $34.0^{+0.7+0.3+0.3}_{-0.6-0.2-0.2}$\\
$\bar B_s^0\to K^+_0\rho^-$& $24.5^{+11.9+9.2+1.8+1.6}_{-9.7-7.8-3.0-1.6}$\\
\hline \hline
\end{tabular}
\end{center}
\end{table}
\begin{figure}[t,b]
\begin{center}
\includegraphics[scale=0.7]{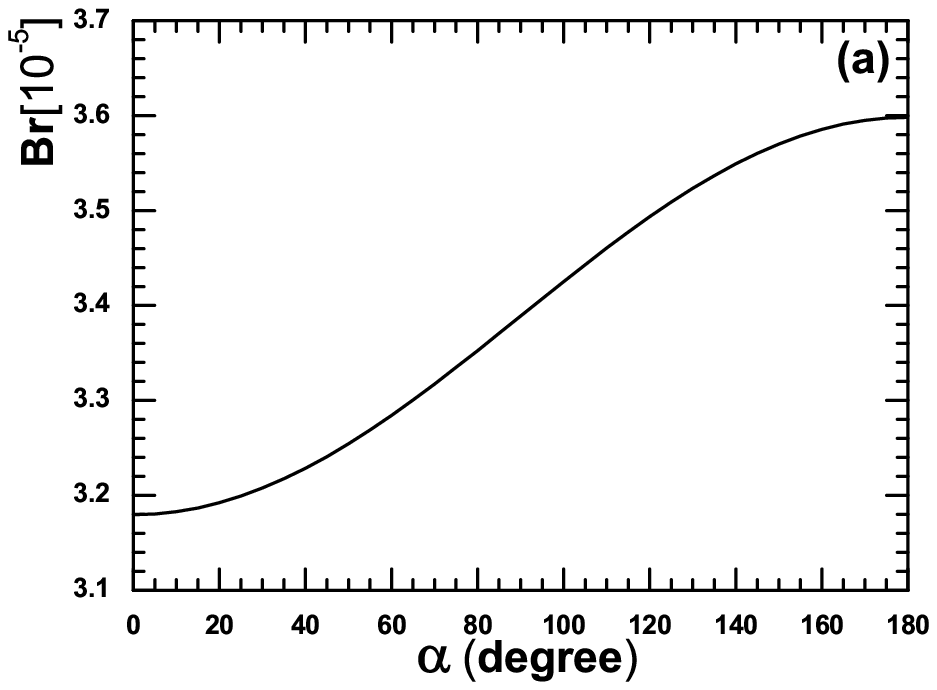}
\includegraphics[scale=0.7]{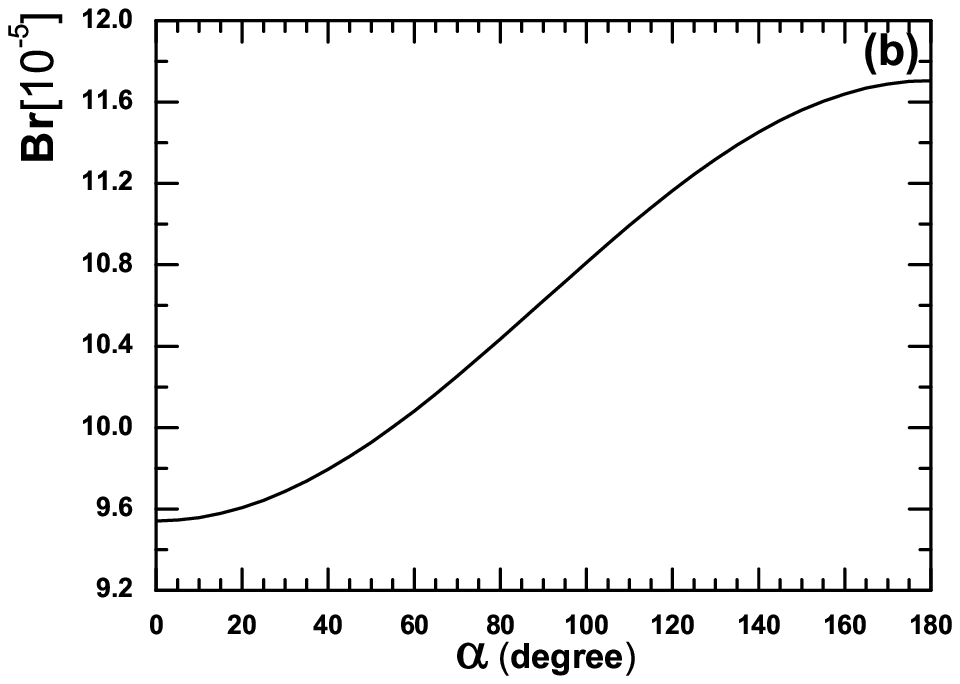}
\vspace{0.3cm} \caption{The dependence of the branching ratio for
$\bar B^0_s\to K^{*+}_0(1430)\rho^-$ on the
Cabibbo-Kobayashi-Maskawa angle $\alpha$. The left (right) panel is plotted in scenario I (II).}\label{fig3}
\end{center}
\end{figure}
\begin{table}
\caption{ Decay amplitudes for decays $\bar B^0_s\to K^{*+}_0(1430)\rho^-, K^{*0}_0(1430)\rho^0$ ($\times 10^{-2} \mbox {GeV}^3$).}
\begin{center}
\begin{tabular}{cc|c|c|c|c|c|c|c}
\hline \hline  &&$F^T_{eK^*_0}$&$F_{eK^*_0}$ & $M^T_{eK^*_0}$ &$M_{eK^*_0}$& $M_{aK^*_0}$ &$F_{aK^*_0}$\\
\hline
$\bar B^0_s\to K^{*0}_0(1430)\rho^0$ (SI) &   &-22.4&4.9&$-11.7+8.2i$&$-0.15+0.24i$&$-0.14+0.11i$ &$-4.1-2.9i$\\
$\bar B^0_s\to K^{*+}_0(1430)\rho^-$ (SI) &   &203&-8.6&$6.5-5.2i$&$0.08-0.28i$&$0.16-0.09i$ &$5.3+4.4i$\\
$\bar B^0_s\to K^{*0}_0(1430)\rho^0$ (SII) &   &27.6&-7.8&$0.7+5.9i$&$0.36-0.20i$&$0.40+0.20i$ &$3.1+8.1i$\\
$\bar B^0_s\to K^{*+}_0(1430)\rho^-$ (SII) &   &-371&14.3&$-0.04-4.6i$&$0.58+0.41i$&$-0.58-0.28i$ &$-4.1-11.5i$\\
\hline \hline
\end{tabular}\label{amp}
\end{center}
\end{table}
In Table II, we list the values of the factorizable and nonfactorizable amplitudes
from the emission and annihilation topology diagrams of the decays $\bar B^0_s\to K^{*0}_0(1430)\rho^0$ and $\bar B^0_s\to K^{*+}_0(1430)\rho^-$.
$F_{e(a)K^*_0}$ and $M_{e(a)K^*_0}$
are the $\rho$ meson emission (annihilation) factorizable
contributions and nonfactorizable contributions from penguin operators respectively. The upper label $T$ denotes the
contributions from tree operators. For the decay $\bar B^0_s\to K^{*0}_0(1430)\rho^0$, there are not diagrams obtained by exchanging
the position of $K^{*0}_0$ and $\rho^0$ in Fig.1, so there are not contributions from $F_{e(a)\rho}$ and $M_{e(a)\rho}$. It is same for
the decay $\bar B^0_s\to K^{*+}_0(1430)\rho^-$.
From Table II, one can find that because of the large Wilson coefficients $a_1=C_2+C_1/3$,
the tree-dominated decay channel $\bar B^0_s\to K^{*+}_0(1430)\rho^-$ receives a large branching ratio value in both scenarios
compared with $\bar B^0_s\to K^{*0}_0(1430)\rho^0$.

The dependence of the branching ratios for the decays $\bar B^0_s\to K^{*0}_0(1430)\rho^0, K^{*0}_0(1430)\omega, K^{*+}_0(1430)\rho^-$
on the Cabibbo-Kobayashi-Maskawa angle $\alpha$
is displayed in Fig.2 and Fig.3. The branching ratios of the $K^{*0}_0(1430)\rho^0$ and $K^{*+}_0(1430)\rho^-$ modes increase with $\alpha$, while
that of the $K^{*0}_0(1430)\omega$ mode decreases with $\alpha$. The values of $\cos\delta$ [shown in Eq.(\ref{brann})]
for the decay modes $\bar B^0_s\to K^{*0}_0(1430)\rho^0, K^{*+}_0(1430)\rho^-$ are opposite in sign with that of $\bar B^0_s\to K^{*0}_0(1430)\omega$ ,
and as a result the behaviors of the branching ratios with the Cabibbo-Kobayashi-Maskawa angle $\alpha$ for the former are very different
with that of the latter. We can also find that the branching ratio of the decay
$\bar B^0_s\to K^{*+}_0(1430)\rho^-$ is insensitive to the variation of $\alpha$ in scenario I. For the decay
$\bar B^0_s\to K^{*0}_0(1430)\phi$, there are only penguin operator contributions in this channel, so its branching ratio has
no relation with the angle $\alpha$ at the leading order.

Now, we turn to the evaluations of the direct CP-violating asymmetries of
the considered decays in the PQCD approach. The direct CP-violating asymmetry can be defined as
\be
\acp^{dir}=\frac{ |\overline{\cal M}|^2-|{\cal M}|^2 }{
 |{\cal M}|^2+|\overline{\cal M}|^2}=\frac{2z\sin\alpha\sin\delta}
{1+2z\cos\alpha\cos\delta+z^2}\;. \label{dirdifine}
\en
Here the ratio $z$ and the strong phase $\delta$ are calculable in PQCD approach, so it is easy to find the numerical values
of $\acp^{dir}$ (in unit of $10^{-2}$) by using the input parameters listed in the previous for the considered decays
in two scenarios:
\be
\acp^{dir}(\bar B^0_s\to K^{*0}_0(1430)\omega)=-24.1^{+0.0+2.7+0.6}_{-0.0-2.5-0.2}, \mbox{ scenario I},\\
\acp^{dir}(\bar B^0_s\to K^{*0}_0(1430)\rho^0)=26.6^{+0.0+2.5+0.3}_{-0.0-2.5-0.5},\mbox{ scenario I},\\
\acp^{dir}(\bar B^0_s\to K^{*+}_0(1430)\rho^-)=7.7^{+0.0+0.2+0.2}_{-0.0-0.3-0.2},\mbox{ scenario I},\\
\acp^{dir}(\bar B^0_s\to K^{*0}_0(1430)\omega)=-86.7^{+0.1+7.1+1.3}_{-0.1-5.3-2.8},\mbox{ scenario II},\\
\acp^{dir}(\bar B^0_s\to K^{*0}_0(1430)\rho^0)=84.5^{+0.1+4.9+1.0}_{-0.1-6.3-3.8},\mbox{ scenario II},\\
\acp^{dir}(\bar B^0_s\to K^{*+}_0(1430)\rho^-)=12.6^{+0.0+0.2+0.8}_{-0.0-0.2-0.6},\mbox{ scenario II},
\en
where the uncertainties are mainly from the decay constant,
the Gegenbauer moments $B_1$ and $B_3$ of the scalar meson $K^*_0$. Compared with the values
of the branching ratios, we can find that if the direct CP-violating asymmetries are sensitive to some
parameters, while the branching ratios are insensitive to them, for example, the decay constant of $K^*_0$.
For the decays
$\bar B^0_s\to K^{*0}_0(1430)\omega, K^{*0}_0(1430)\rho^0$, their direct CP-violating asymmetries in scenario II are more than
3 times than those in scenario I. In both scenarios, the direct CP-violating asymmetries of these two decay channels are
close to each other in size, while they are opposite in sign. The reason for this is the following. The mesons $\rho^0, \omega$ have very similar mass,
decay constant, and distribution amplitude, only the opposite sign of $d\bar d$ in their quark components, and the difference will appear in
penguin operators. From our numerical results, we can find that the contributions from tree operators for these two channels (denoted as $T_{K^{*0}_0\rho^0}$
and $T_{K^{*0}_0\omega})$ are really very close, and those from penguin operators for these two channels (denoted as $P_{K^{*0}_0\rho^0}$
and $P_{K^{*0}_0\omega}$) are opposite in sign. Furthermore, the real parts of $P_{K^{*0}_0\rho^0}$ and $P_{K^{*0}_0\omega}$ in each scenario have large
differences in size.
\be
T_{K^{*0}_0\rho^0}&=&(-34.1+i8.2)\times10^{-2}, P_{K^{*0}_0\rho^0}=(0.49-i2.5)\times10^{-2},\\
T_{K^{*0}_0\omega}&=&(-31.8+i7.6)\times10^{-2}, P_{K^{*0}_0\omega}=(-3.7+i2.8)\times10^{-2}, \mbox{ scenario I},\\
T_{K^{*0}_0\rho^0}&=&(28.3+i5.9)\times10^{-2}, P_{K^{*0}_0\rho^0}=(-3.9+i8.1)\times10^{-2},\\
T_{K^{*0}_0\omega}&=&(26.4+i5.5)\times10^{-2}, P_{K^{*0}_0\omega}=(8.1-i7.2)\times10^{-2}, \mbox{ scenario II}.
\en
These values can explain why the two channels have similar CP-violating asymmetry in size (certainly, their branching ratios are also similar for
the same reason). Using the
upper results, we can calculate $\sin\delta$ [shown in Eq.(\ref{dirdifine})] in two scenarios:
\be
\sin\delta_{K^{*0}_0\rho^0}&=&0.91, \sin\delta_{K^{*0}_0\omega}=-0.40, \mbox{ scenario I},\\
\sin\delta_{K^{*0}_0\rho^0}&=&0.97, \sin\delta_{K^{*0}_0\omega}=-0.80, \mbox{ scenario II}.
\en
These values can explain why the CP-violating asymmetries of these two decays have opposite signs.

The direct CP-violating asymmetry of $\bar B^0_s\to K^{*+}_0(1430)\rho^-$ is the smallest in these decays, about $10\%$,
but its branching ratio is the largest one,
about $3.4\times 10^{-5}$ in scenario I, even at the order of $10^{-4}$ in scenario II. So this channel might be easily measured at LHC-b experiments.
\begin{figure}[t,b]
\begin{center}
\includegraphics[scale=0.7]{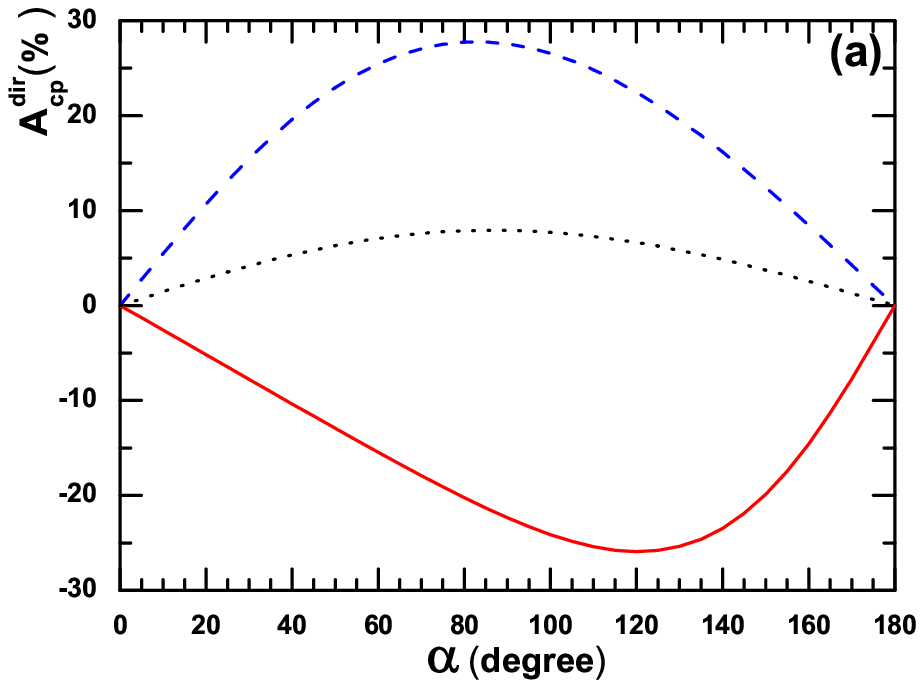}
\includegraphics[scale=0.7]{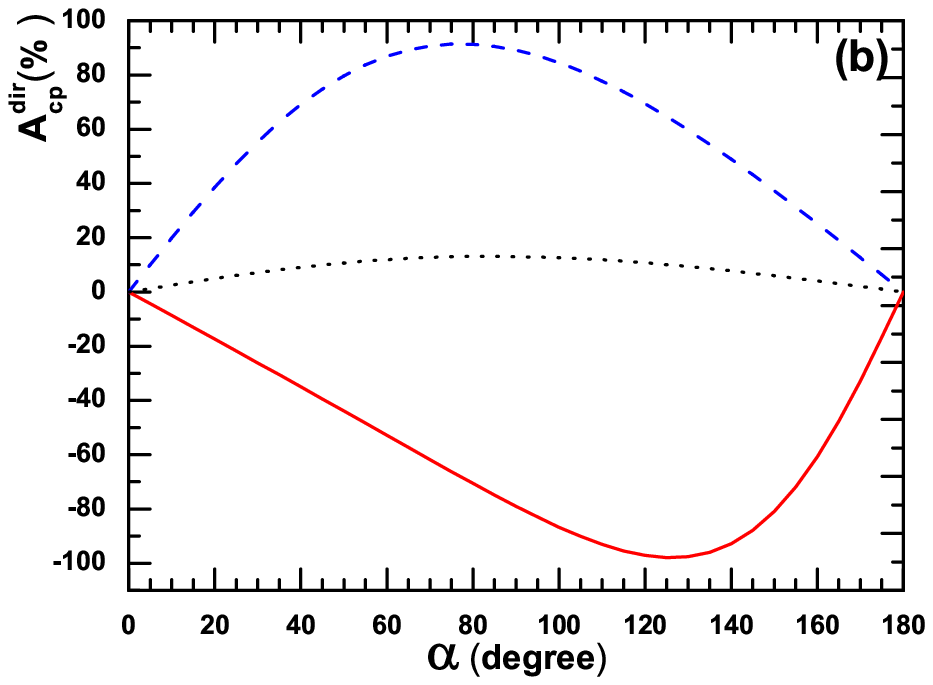}
\vspace{0.3cm} \caption{The dependence of the direct CP asymmetries for
$\bar B^0_s\to K^{*0}_0(1430)\omega$ (solid curve), $\bar B^0_s\to K^{*+}_0(1430)\rho^-$ (dotted curve), $\bar B^0_s\to K^{*0}_0(1430)\rho^0$ (dashed curve) on the
Cabibbo-Kobayashi-Maskawa angle $\alpha$. The left (right) panel is plotted in scenario I (II)}\label{fig2}
\end{center}
\end{figure}
From Fig.4(a) and 4(b), one can see that though the direct CP asymmetry values for each decay in two scenarios are very different
in size, they
have similar trends depending on the Cabibbo-Kobayashi-Maskawa angle $\alpha$. As for the decay $\bar B^0_s\to K^{*0}_0(1430)\phi$, there is no tree contribution at the
leading order, so the direct CP-violating asymmetry is naturally zero.


\section{Conclusion}\label{summary}

In this paper, we calculate the branching ratios and the CP-violating
asymmetries of decays $\bar B_s^0\to K^*_0(1430)\rho(\omega,\phi)$
in the PQCD factorization approach.
Using the decay constants and light-cone distribution amplitudes
derived from QCD sum-rule method, we find that
\begin{itemize}
\item
We predict the form factor $A^{\bar B^0_s\to \phi}_0(q^2=0)=0.29^{+0.05+0.01}_{-0.04-0.01}$
for $\omega_{b_s}=0.5\pm0.05$ and the Gegenbauer moment $a_{2\phi}=0.18\pm0.08$, which agrees well with the values as
calculated by many approaches and disagrees with the value $A^{\bar B^0_s\to\phi}_0=0.474$ obtained by the light-cone sum-rule method. The
discrepancy can be clarified by the LHC-b experiments.
The form factors of $\bar B^0_s\to K^*_0(q^2=0)$ in two scenarios are given as
\be
F^{\bar B^0_s\to K^*_0}_0(q^2=0)&=&-0.30^{+0.03+0.01+0.01}_{-0.03-0.01-0.01}, \quad\mbox{ scenario I},
\\F^{\bar B^0_s\to K^*_0}_0(q^2=0)&=&0.56^{+0.05+0.03+0.04}_{-0.07-0.04-0.05}, \quad\;\;\;\mbox{ scenario II},
\en
where the uncertainties are from the decay constant,
the Gegenbauer moments $B_1$ and $B_3$ of the scalar meson $K^*_0$.
\item
Because of the large Wilson coefficients $a_1=C_2+C_1/3$,
the branching ratios of $\bar B^0_s\to K^{*+}_0(1430)\rho^-$ are much larger than those of the other three decays in both scenarios
and arrive at a few times $10^{-5}$ in scenario I, even at the $10^{-4}$ order in scenario II, while its direct CP-violating asymmetry is the smallest one, around
$10\%$. The values for this channel might be measured by the current LHC-b experiments.
\item
For the decays $\bar B^0_s\to K^{*0}_0(1430)\omega, K^{*0}_0(1430)\rho^0$, their direct CP-violating asymmetries are large, but it
might be difficult to measure them, because their branching ratios are small and less than (or near) $10^{-6}$ in both scenarios.
\item
The values of $\cos\delta$ for the decays $\bar B^0_s\to K^{*0}_0(1430)\rho^0, K^{*+}_0(1430)\rho^-$
are opposite in sign with that for $\bar B^0_s\to K^{*0}_0(1430)\omega$;
as a result, the behaviors of the branching ratios of the former with Cabibbo-Kobayashi-Maskawa angle $\alpha$ are very different with that of the latter.
Because the values of $\sin\delta$ are opposite in sign,
their direct CP-violating asymmetries of the former have an opposite sign with that of the latter. Here $\delta$ is the relative strong phase angle between
the tree and the penguin amplitudes.
\item
Because the mesons $\rho^0, \omega$ have very similar mass,
decay constant, distribution amplitude, only opposite sign of $d\bar d$ in their quark components, and this difference only appears in the
penguin operators; so these two tree document decays $\bar B^0_s\to K^{*0}_0(1430)\rho^0$ and $\bar B^0_s\to K^{*0}_0(1430)\omega$ should have similar
branching ratios and CP-violating asymmetries.
\item
As for the decay $\bar B^s_0\to K^{*0}_0(1430)\phi$, though there exist large differences between the two scenarios,  the predicted branching ratios
are small and a few times $10^{-7}$ in both scenarios. There is no tree contribution at the
leading order, so the direct CP-violating asymmetry is naturally zero.
\end{itemize}

\section*{Acknowledgment}
This work is partly supported by the National Natural Science
Foundation of China under Grant No. 11047158, and by Foundation of
Henan University of Technology under Grant No.150374. The author
would like to thank Cai-Dian L\"u for helpful discussions.

\end{document}